\documentclass[prb,twocolumn,showpacs, amsmath, amssymb]{revtex4}
\allowdisplaybreaks
\usepackage{epsfig}
\usepackage{graphicx}

\newcommand{\beq} {\begin{equation}}
\newcommand{\eeq} {\end{equation}}
\newcommand{\beqa} {\begin{eqnarray}}
\newcommand{\eeqa} {\end{eqnarray}}

\begin{document}
\title{A Selective Advantage for Conservative Viruses}
\author{Yisroel Brumer and Eugene I. Shakhnovich}
\affiliation{Harvard University, 12 Oxford Street, Cambridge, Massachusetts 02138}
\date{\today}
\begin{abstract}
In this letter we study the full semi-conservative treatment
of a model for the co-evolution of a virus and an adaptive immune system. Regions
of viability are calculated for both conservatively and semi-conservatively 
replicating 
viruses interacting with a realistic semi-conservatively replicating immune system. 
The conservative virus is found to have a selective advantage in the form of
an ability to survive in regions with a wider range of mutation rates than
its semi-conservative counterpart. 
This may help explain the existence of 
a rich range of viruses with conservatively replicating genomes, a trait which is found nowhere else
in nature.
\end{abstract}

\maketitle

DNA is often called the molecule of life. 
The vast majority of organisms in nature store their genetic information in
the form of double stranded DNA, which provides a number of benefits over 
its close relative and likely predecessor, RNA \cite{Voet}.  
These benefits include a predictable secondary structure and a resistance to
auto-catalytic cleavage and hence a longer half-life. However, viruses stand
out as a notable 
exception to the ``DNA world'', employing a variety of methods for genetic storage 
including
single stranded RNA (e.g. tobacco mosaic virus), double stranded RNA 
(such as reovirus), linear and circular single stranded DNA (including parvovirus and
bacteriophage $\phi$X174, respectively) and a variety of double stranded DNA
types (examples of which include bacteriophage T4, polyoma virus and poxvirus). 
While double stranded DNA
replicates semi-conservatively, many of these viruses
replicate conservatively, wherein multiple copies of
a single strand are made, and the original strand is conserved.  This suggests
that either the benefits of DNA must be less for viral species 
(possibly due to a shorter genome length or unusual life cycles) and/or a selective advantage must exist for 
conservatively replicating viruses. In this paper we use the the quasispecies
model to search for such an advantage.

Introduced by Eigen in 1971,  the quasispecies
model \cite{Eigen1, Eigen} has been used to study various 
characteristics of conservatively replicating
organisms ranging from equilibrium data to punctuated evolution\cite{landscape,
 timedep, timedep2, Tarazona, Galluccio, Schuster, Peliti, Altmeyer, Campos, 
Alves, Wilke, Krug}. The model consists of a population of independently replicating genomes 
$ \phi $, each of which is made up of a set of ``letters'' $s_1s_2\cdots s_n$
chosen from an ``alphabet'' of size S. S is usually chosen to be two for 
simplicity or four, as in this paper, to model the nucleotides A, C, T and G. 
Each possible genome is assigned a fitness that dictates
its fecundity. This mapping of fitness to genotype can be represented by a 
unique ``fitness landscape''. The process of replication includes a probability of
point mutation per base pair
$ \epsilon $ that is generally assumed to be genome-independent. By associating
phenotype with genotype and assuming a first-order dependence of the growth
rate on concentration, a set of differential equations can be solved to 
describe the competition between various genotypes \cite{Eigen1, Eigen}.
Although the model incorporates numerous approximations, it is well suited to 
describing small RNA genomes and viruses and many of its predictions
have been experimentally verified. One of the major successes of the model
lies in recent work on novel anti-viral therapies \cite{Crotty,
Loeb}.

The quasispecies model has recently been extended to the coevolution of hosts and parasites and the particular case of an adaptive immune system interacting with a virus\cite{Kamp, Kamp2}. Viruses
make detrimental use of host biochemical processes while the immune system
expends enormous effort to keep viral concentrations as low as possible. 
As the immune system develops new defenses, the virus must adapt to defeat them.
The immune system must then evolve to destroy the newly resistant
strains, and a non-linear co-evolving feedback loop is created.
To
model this behavior, the immune system and virus are both assumed to evolve on 
a single fitness peak landscape, where the fitness of all genomes are 
equal with the sole exception of a single master sequence of far greater fitness, or
\begin{equation}
A(\phi) = \left\{ \begin{array}{cc}
             \eta & \phi \neq \phi_0 \\
             \sigma \gg \eta & \phi = \phi_0
             \end{array}
             \right. ,
\end{equation}
where $ A(\phi) $ represents the fitness of genome $ \phi $ and $ \{\sigma, 
\eta\} \equiv \{\sigma_v, \eta_v\} $ for the virus and  $\{\sigma, \eta\} \equiv 
\{\sigma_{is}, \eta_{is}\} $ for the immune system. This landscape is dynamic
in that the fitness peak is allowed to shift from one genome to another
at specified intervals. To model the inter-species interaction, the master 
sequence
for the immune system is
assumed to impose a death rate $ \delta $ on the corresponding viral sequence.
When this coincides with the viral master sequence, the viral fitness peak
shifts out of self-preservation. The new viral master
sequence regenerates on a time scale $ \tau_v $, defined as the time
required for the new master sequence to outnumber the old. At this point, the
immune fitness peak shifts to match the viral master sequence and will regenerate
on a similarly defined timescale $ \tau_{is} $ after which the viral peak
shifts again. These steps are iterated so that the virus
traverses genome space with the immune system in hot pursuit. Using recent results on the 
dynamics of a quasispecies on time-dependent landscapes \cite{timedep2}, 
Kamp and Bornholdt \cite{Kamp, Kamp2} found expressions for the long-term 
survival of a {\em conservatively}
replicating virus and immune system \cite{Kamp, Kamp2} by considering the behavior of each species on a dynamic fitness landscape. In essence, every time the fitness
peak shifts, the concentration of the master sequence drops dramatically and 
begins to regrow, while the large concentration of the old master sequence
drains away. 
Rigorously,
the dynamics of a quasispecies on a single fitness peak landscape follow a set of differential equations
defining the 
evolution of the various
genomes in terms of their Hamming distance $ HD(\phi,\phi') $. This is defined
as the smallest number of point mutations that lead from $\phi$ to $\phi'$. Utilizing the radial symmetry of the fitness landscape, the quasispecies equations 
become
\begin{eqnarray}\nonumber
&&\frac{dw_{l,x}}{dt} = \\[0.25in]\nonumber 
&&\sum_{l' = 0}^{l} \frac{(n_x - l')!}{(n_x - l)!}A_x(l')(\frac{\epsilon_x}{S-1})^{l-l'}(1-\epsilon_x)^{n_x - (l - l')}w_{l',x} \\[0.25in]
&& - f_x(t)w_{l,x},
\end{eqnarray}
where $x \in \{v, is\}, w_l $ is the concentration of sequences of Hamming distance $l$ from the
master sequence, $ f_{is}(t) = \sum_{l}A_{is}(l)w_{l, is} = (\sigma_{is} - \eta_{is})w_{0, is} + \eta_{is} $ for
the immune system, $f_v(t) = \delta $ for the  viral sequence that coincides
with the immune master sequence and $f_v(t) = 0 $ otherwise. 
$\epsilon $ represents the point mutation probability (more complex mutations
such as insertions and deletions, as well as the possibility of recombination,
are ignored). 

Building on recent work regarding the dynamics of a quasispecies on mobile
fitness landscapes \cite{timedep2}, a number of qualitative features have come 
to light. Besides the standard (albeit modified by the dynamic nature of the 
landscape) error catastrophe at high $ \epsilon_v $,
the virus is only viable above a given minimum mutation rate, below which it
is unable to keep up with the moving landscape (dubbed the ``adaptability'' 
catastrophe). At these low mutation rates, each shift of the landscape is
followed by a period of time wherein the new master sequence attempts to build
up an equilibrium distribution. However, before the new master sequence can 
rebuild to the levels reached by the old master before the shift, the fitness 
peak moves again. As this process repeats, any initial quasispecies will 
disappear and genomes become stochastically distributed.
Further, an optimal mutation rate for the immune system can be found that 
minimizes the range
of viral mutation rates that allow persistence of the viral quasispecies. 
This optimal
rate has been determined and found to be independent of the parameters
of the model and the properties of the viral species as well as comparing 
admirably with the rates of somatic hypermutation in B-cells \cite{Kamp}.
The agreement suggests that this model, although approximate in nature,
captures the robust features of the co-evolution of viruses with an adaptive immune system.

\begin{figure}[htb]
\includegraphics[width=1\linewidth]{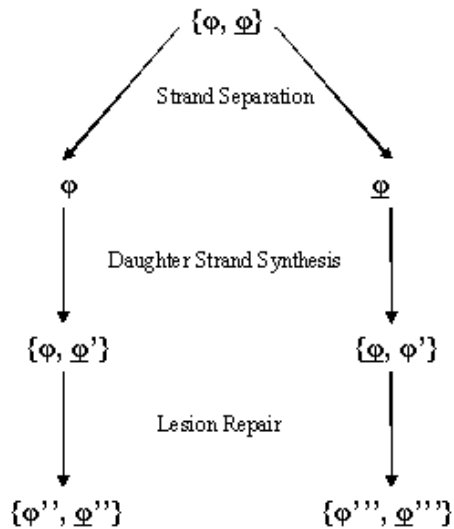}
\caption{A schematic model of DNA replication. Adapted from \cite{Manny}.}
\end{figure}

Although the host-parasite model has yielded impressive successes, work 
has been restricted to conservatively replicating 
systems, which differ greatly from the true semi-conservative systems that
dominate nature. In a conservative system, multiple, possibly error prone,
copies of an original strand are produced without harming or changing the 
original strand. Thus, the original quasispecies model is ideally suited for the 
study of RNA viruses or {\it in-vitro} RNA evolution experiments. 
Semi-conservative replication follows a different route 
shown schematically in Fig. 1. DNA exists as a tightly bound double helix
structure, where each strand $ \phi $ is connected to a complementary strand
$\underline{\phi}$, where $ \underline{\phi} $ represents the complement
of strand $ \phi $, and can be written as $ \underline{s}_1\underline{s}_2
\cdots \underline{s}_n $ where we assign the nucleotides $ A \equiv 1, 
G \equiv 2, T \equiv 3, C \equiv 4 $ and $\underline{s}_i = (s_i + S/2) \mod S $.
In order to undergo replication, the double helix unzips to free two single strands $\phi$ and 
$\underline{\phi} $. Each of these is replicated to produce an error-prone
complement, yielding $ \{\phi, \underline{\phi}'\} $ and $ \{\underline{\phi},
\phi'\} $. Each error must result in a base mismatch, which can be recognized and
selectively repaired by enzymes in the cell.  These enzymes can recognize the
new strand and ensure that the mismatch is repaired by replacing the new mismatched base,  keeping the effective
error rate $ \epsilon $ low. In the final stage, the strands become 
indistinguishable and various maintenance enzymes repair the remaining
errors with a 50\% probability of correcting the mismatch in either strand.
Thus, the final result are two new pairs, each consisting of two new strands,
$ \{\phi'', \underline{\phi}''\} $ and $\{\phi''', \underline{\phi}'''\} $. Recent work 
by Tannenbaum {\it et. al} \cite{Manny} has extended the quasispecies model to the case of 
semi-conservative replication, which was found to display significantly different 
behavior in the infinite time limit on a static landscape. 

To properly treat a semi-conservative quasispecies model on a single fitness 
landscape, ignoring back mutations, Eqn. (2) must be
recast as \cite{Manny}
\begin{eqnarray}\nonumber
&&\frac{dw_{l, x}}{dt} = 2\sum_{l' = 0}^l A_x(l')(\frac{\epsilon_x/2}{S-1})^{l-l'}(1-\frac{\epsilon_x}{2})^{n_x - l - l'}w_{l',x} \\[0.25in] 
&&- (A_x(l) + f(t))w_{l,x},
\end{eqnarray}
where$ f(t) $ is defined above. Here, we examine the dynamics of the
semi-conservative equations within the confines of Kamp and Bornholdt's model 
of co-evolution. This study will focus on the dynamical aspects of Eqn. (3), as
opposed to the equilibrium effects studied by Tannenbaum {\it et. al.}\cite{Manny}.
Following Kamp and Bornholdt, the condition for the viability of the 
quasispecies is 
\begin{equation}
\kappa_x \equiv \frac{w_{1,x}(\tau)}{n_x e^{\eta_x \tau} w_{0,x}(0)} \geq 1,
\end{equation}
where $\kappa $ represents the ratio of master sequence concentrations at the
beginning and end of an
entire cycle of landscape shifts in an unconstrained system compared to
the equivalent growth of a random sequence far from the peak. Obviously,
if the master sequence outgrows the random sequence over this period, even
with the concentration losses incurred by the peak shift, it will survive 
for all times. If the master sequence is outgrown by the random sequence, i.e.
if $\kappa < 1$, the master sequence will diminish and disappear at long times.

After a fair bit of work, the condition for viability of the immune genome can be expressed as\cite{Brumer}
\begin{eqnarray} \nonumber
&& \kappa_{is} = \left(\frac{\sigma_{is} \epsilon_{is} (1 - \epsilon_{is}/2)^{n_{is} - 1}}{(S-1)(\sigma_{is} - \eta_{is})(2(1- \epsilon_{is}/2)^{n_{is}} - 1)}\right)  \\[0.25in] \nonumber
&& \times (e^{(2\sigma_{is}(1-\epsilon_{is}/2)^{n_{is}} - \sigma_{is} - \eta_{is})\tau} - \\ [0.25in]
&& e^{(2\eta_{is}(1 - \epsilon_{is}/2)^{n_{is}} - 2\eta_{is})\tau}) \geq 1\\ [0.25in]&& \tau = \tau_{is} + \tau_{v} \\[0.25in]
&& \tau_{is} = -\frac{\ln \left( \frac{(1-\epsilon_{is}/2)^{n_{is}}\epsilon_{is}}
{(2(1-\epsilon_{is}/2)^{n_{is}} - 1)(S-1)}\right)}
{(2(1-\epsilon_{is}/2)^{n_{is}} - 1)(\sigma_{is} - \eta_{is})}
\end{eqnarray}
and, for a semi-conservative viral species,
\begin{eqnarray} \nonumber
&& \kappa_{v} = \left(\frac{\sigma_{v} \epsilon_{v} (1 - \epsilon_{v}/2)^{n_{v} - 1}}{(S-1)(\sigma_{v} - \eta_{v})(2(1- \epsilon_{v}/2)^{n_{v}} - 1)}\right)  \\[0.25in] \nonumber
&& \times (e^{(2\sigma_{v}(1-\epsilon_{v}/2)^{n_v} - \sigma_{v} - \eta_v)\tau} - \\ \nonumber
\\  && e^{(2\eta_{v}(1 - \epsilon_{v}/2)^{n_{v}} - 2\eta_{v})\tau}) \geq1 \\[0.25in]
&& \tau_{v} = -\frac{\ln \left( \frac{(1-\epsilon_{v}/2)^{n_v}\epsilon_{v}}
{(2(1-\epsilon_v/2)^{n_{v}} - 1)(S-1)}\right)}
{(2(1-\epsilon_v/2)^{n_{v}} - 1)(\sigma_{v} - \eta_{v}) + \delta}. \label{}
\end{eqnarray}
A conservatively replicating virus interacting with a semi-conservative
immune system will follow the behavior described by the conservative model
of Kamp and Bornholdt,
\begin{equation}
\kappa_v = \left( \frac{(e^{(q_v^n\sigma_v -\eta_v)\tau} - e^{(q_v^{n_v}\eta_v -\eta_v)  \tau})(1 - q_v)\sigma_v}{(S-1)(\sigma_v - \eta_v)q_v}\right),
\end{equation}
where $ q = 1- \epsilon $ represents the replicative fidelty per base pair, 
$\tau = \tau_{is} + \tau_v $ as always, and
\begin{eqnarray}
&& \tau_v = -\frac{\ln(\frac{1-q_v}{S-1})}{q_v^n(\sigma_v - \eta_v) + \delta}.
\end{eqnarray}
In contrast to previous work, $\tau_{is}$ is described by the semi-conservative
result described by Eqn. (7).
\begin{figure}[htb]
\includegraphics[width=1.0\linewidth]{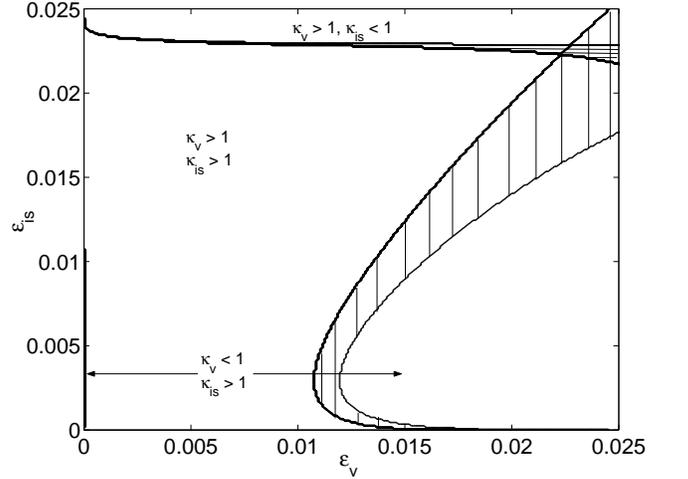}
\caption{Regions of viability for a co-evolving host-parasite system. The
contours for both semi-conservative and conservative virus interacting with 
a conservative immune system are shown. The vertical lines represent the 
region where only the conservative virus is viable, while the horizontal
lines represent the region where the immune system is only viable for a
conservative virus. $n_{is} =n_v = 50, \sigma_{is} = \sigma_v = 100, \eta_{is} = \eta_v = 1, \delta = 200 $.
}
\end{figure}
In Fig. 2 we plot the regions of viability defined by $\kappa_{v, is} \geq 1$ with a particular set of parameters for a semi-conservative immune receptor 
interacting with both a conservative and semi-conservative virus.
It is immediately clear that the conservative
virus can survive under a wider range of conditions than the semi-conservative,
thus obtaining a selective evolutionary advantage. While the immune system is
slightly less robust for the semi-conservative virus, this effect is small
and lies near the region where the immune system becomes unviable, and hence
should have little effect on real systems. Although this behavior
is dependent on the parameters of the model, the qualitative trend was robust
for the vast majority of biologically reasonable parameter choices. 
Further, the 
increased viability lies both near and far from the optimal immune mutation rate, important since a 
population of viruses in nature is expected to interact with a range
of immune system properties, and the behavior away from the optimum
should play an important role in the evolution of the system. Thus, we have demonstrated the existence of a clear selective advantage for conservative 
replication in viral species.

We conclude with a cautionary note. 
One must always take great care in extracting grand predictions from
simplified models such as this one. In nature, a myriad of evolutionary
pressures battle for dominance and it is often difficult to pinpoint the
selective advantage for a given trait. However, the model presented here
likely captures many robust features of the evolution, and suggests a possible
explanation for the success of conservative viruses in nature.

\begin{acknowledgments}
The authors gratefully acknowledge useful discussions with Emmanuel Tannenbaum
and Brian Dominy.
\end{acknowledgments}

\bibliography{ConsImmunePRL.bib}
\end{document}